# High-pressure synthesis of superconducting $Nb_{1-x}B_2$ (x = 0 – 0.48) with the maximum $T_c$ = 9.2 K

Ayako Yamamoto[1,*], Chigaku Takao[1,2], Takahiko Masui[1,3], Mitsuru Izumi[2] and Setsuko Tajima[1]

1 Superconductivity Research Laboratory, International Superconductivity Technology Center, 1-10-13 Shinonome, Koto-ku, Tokyo 135-0062, Japan

2 Tokyo University of Mercantile Marine, 1-6-2 Etchujima, Koto-ku, Tokyo 135-8533, Japan

3 Domestic Research Fellow, Japan Society for the Promotion of Science, Kawaguchi, Japan

ABSTRACT

Superconductivity with $T_c$ above 9 K was found in metal-deficient $NbB_2$ prepared under 5 GPa, while no clear superconductivity was observed down to 3 K in stoichiometric $NbB_2$. The superconductivity was observed above x = 0.04 in $Nb_{1-x}B_2$. and the lattice parameters also changed abruptly at x = 0.04. As x increased, the transition temperature $T_c$ slightly rose and fell with the maximum value of 9.2 K at x = 0.24 for the samples sintered at 5 GPa and 1200 ˚C. The $T_c$–value changed in the range from 7 K to 9 K, depending on the sintering pressure. A series of $Ta_{1-x}B_2$ (0 x 0.24) was also synthesized under high pressure to examine a special effect of high-pressure synthesis.

1. Introduction

Since the discovery of superconductivity at $T_c$ = 39 K in MgB$_2$ [1], the physical properties of metal borides have been reinvestigated. In particular, the superconductivity of transition-metal diborides with an AlB$_2$-type structure has been extensively studied from both the theoretical [2–4] and experimental [5–7] viewpoints. Comparison of physical properties of various transition metal diborides is not only interesting but also important to understand the mechanism of superconductivity in MgB$_2$.

Physical and structural properties of MB$_2$ (M: transition metals) have been studied from the 1950's for various transition metals such as Sc, Ti, V, Y, Zr, Nb, Hf, and Ta [8–10]. In 1970, Cooper *et al.* reported superconductivity in NbB$_2$, MoB$_2$ and the substituted compounds [11]. They found that stoichiometric NbB$_2$ and MoB$_2$ did not show superconductivity but the samples with B-rich compositions such as NbB$_{2.5}$ and MoB$_{2.5}$ did with $T_c$ of 6.4 K and 8.1 K, respectively. They also reported that a partial substitution of Y$^{3+}$ or Hf$^{4+}$ for Nb$^{5+}$ or Mo$^{6+}$ was effective for raising the $T_c$. In those days, the results were discussed based on so-called Matthias's rule [12], which states that superconductivity is enhanced when an averaged valence deviate from integer. On the other hand, superconductivity in stoichiometric NbB$_2$ was reported by Leyarovska [13] in 1957, though the $T_c$ was only 0.62 K. In their study, superconductivity was not observed in MB$_2$ (M = Ti, Zr, Hf, V, Ta, Cr and Mo) down to 0.42 K [13].

Recently, interest in studies of MB$_2$ has been renewed. Akimitsu *et al.* achieved the superconductivity with a $T_c$ of 5.2 K in NbB$_2$ prepared in a sealed quartz tube [14]. They also pointed out that the B-rich composition led to a higher $T_c$. On the other hand, Gasparov *et al.* [5] and Kaczorowski *et al.* [6] reported that superconductivity was absent in NbB$_2$ down to 2 K. Therefore, it is still controversial whether NbB2 exhibits superconductivity or not. Superconductivity in TaB$_2$ is also being discussed at present. Recently, Kaczorowski *et al.* reported superconductivity with a $T_c$ = 9.5 K in TaB$_2$ [6], whereas Rosner *et al.* did not achieve superconductivity down to 1.5 K [15]. Although the reason for the contradictions has not been made clear yet, it is still worth examining the relationship between the composition and superconductivity in $NbB_2$ and the related diborides.

In general, high-pressure synthesis is a very efficient method of the search for new superconducting materials [16, 17]. A closed system in high-pressure synthesis is effective not only for stabilizing a composition but also for expanding a solid-solution range. As for borides, it is difficult to prepare a highly dense polycrystalline sample when using a conventional solid-state reaction method because the melting temperature is extremely high. In deed, Itoh *et al*. obtained desirable $TaB_2$ samples for density measurements by annealing under high-pressure [18].

In the present study, we demonstrate the superconductivity of $Nb_{1-x}B_2$ prepared under high pressure and discussed the compositional, dependence of structure and superconductivity. Comparing $Nb_{1-x}B_2$ with the other metal diborides, similarities and differences are also discussed.

2. Experimental procedures

Samples were synthesized from Nb, Ta and Mo powder (99.9 %), and B amorphous (97 %). The amorphous B was preheated at 800 ˚C for 12 h in a vacuum ($10^{-5}$ Torr) to remove humidity and other volatile impurities. Mixed powers with the ratio M : B = (1-x) : 2 (M = Nb, Ta and Mo) for various x's were mounted into a sintered BN tube. The samples were heated to 900–1300 ˚C in 10 min and sintered for 0.5–2 h under the pressure of 1–5 GPa using cubic-anvil-type equipment (Try Eng. Co.). They were finally quenched to room temperature within a few seconds.

The samples were characterized by the X-ray power diffraction (XRD) pattern (Mac Science MX18). Lattice parameters were determined using the XRD patterns by the least-squares method. Electron microprobe analysis (EPMA) was performed using a scanning electron microscope (SEM, Hitachi JXM-8600MX) with an accelerated voltage of 15 kV. The DC susceptibility was measured with superconducting quantum interference devices (SQUID, Quantum Design MPMS) in both zero-field and field-cooling modes. The electrical resistivity and thermoelectric power were measured by the conventional four-probe method for bar-shaped samples.

## 3. Results

### 3.1 Synthesis and superconductivity of $Nb_{0.84}B_2$

In order to confirm the superconductivity and to optimize the synthetic condition in the $Nb_{1-x}B_2$ system, we first synthesized two compositions of samples, stoichiometric $NbB_2$ and metal deficient $Nb_{0.84}B_2$. The latter composition is close to that of $NbB_{2.5}$ for which superconductivity was reported [11].

The $NbB_2$ sample sintered at 5 GPa and 1200 °C for 0.5 – 2 h showed relatively broad peaks in the XRD pattern, though no impurity peak was found. Secondary and refracted electron images of the sample by SEM indicated the presence of a small amount of impurity that was identified as Nb by EPMA. Although superconductivity was detected by DC susceptibility below 7 K, the transition was broad and the volume fraction of the superconductivity was less than 10 %, which suggested the inhomogeneity of this sample. On the other hand, the peak width in the XRD of $Nb_{0.84}B_2$ was sharper than that of the stoichiometric sample, and no impurity was detected in either XRD pattern or SEM observation. Moreover, a sharp superconducting transition was observed around at 9 K in susceptibility, which supports better homogeneity. Thus, our experiments began with the nominal composition of $Nb_{0.84}B_2$.

At this stage it is not clear which atom (Nb deficiency or interstitial B) is related to non-stoichiometry in $NbB_2$. Here, we assume it is Nb deficiency in this case because the honeycomb network of B looks very rigid and it may thus be very difficult to find an interstitial space for B. Moreover, it is unlikely that B substitutes for Nb. Details of the non-stoichiometry problem will be discussed in the following section.

First, an appropriate sintering temperature and time were investigated for $Nb_{0.84}B_2$. The temperature was varied from 1000 to 1300 °C with 100 °C intervals under 5 GPa. Unreacted Nb was detected by the XRD pattern only in the samples sintered at 1000 °C. The highest $T_c$ of 9.1 K was obtained in the samples sintered at 1200 °C. The sintering time was varied from 0.5 to 2 h, but this resulted in no change in either the $T_c$ or lattice parameters.

Figure 1 shows the XRD pattern of $Nb_{0.84}B_2$ sintered at 1200 °C for 0.5 h under 5 GPa. All peaks are indexed according to the AlB$_2$-type structure (hexagonal, P6/mmm(no.191)). The indexes, *d*-value and intensity of the pattern are listed in Table 1. The lattice parameters are *a* = 3.0979(4) Å and *c* = 3.321(1) Å, which are close to those for a B-rich sample (NbB$_{2.5}$) with a $T_c$ of 6.4 K reported by Cooper *et al*. [11]. On the other hand, *a* is shorter and *c* is longer than that of stoichiometric sample [11]. Photographic images of secondary and refracted electrons of Nb$_{0.84}$B$_2$ taken by SEM are shown in Fig. 2. These images proved that the sample was of high quality with no impurity. The compositions determined by EPMA were Nb$_{0.76(\pm0.03)}$B$_{2.0}$ and Nb$_{0.68(\pm0.03)}$B$_{2.0}$ for the nominal compositions of Nb$_{0.84}$B$_{2.0}$ and Nb$_{0.76}$B$_{2.0}$, respectively.

The superconductivity of Nb$_{0.84}$B$_2$ was confirmed by DC susceptibility and electrical resistivity, as shown in Fig. 3. A clear diamagnetic signal and a sharp transition are observed below 9.1 K in the susceptibility, and a resistivity drop from at 9.1 K and a zero resistivity at 8.7 K are observed. The absolute value of the resistivity is relatively high and the resistivity curve is *T*-linear over a wide temperature range, which is not similar to the behavior of the other metallic diborides [5]. These results are probably due to a trace of impurity at the grain boundary in the polycrystalline sample. Single crystal is required to study the detailed transport properties.

In order to determine a type of carrier, thermoelectric power of Nb$_{0.76}$B$_{2.0}$ was measured at 297 K. The absolute value of the Seebeck coefficient *S* was –2.0±0.4 V/K. The negative sign of *S* indicates that carriers are electrons. It has been reported that, according to the Hall coefficient measurement, the carriers in stoichiometric NbB$_2$ are also electrons from [10]. Though we cannot simply compare our results with the report because of a difference in the Nb composition, it is reasonable that carriers in both Nb$_{0.84}$B$_2$ and NbB$_2$ are electrons. The electron carriers in Nb$_{0.84}$B$_2$ stand in contrast to hole carriers in MgB$_2$ with a $T_c$ of 39K [19]. This may correspond to the facts that the carriers for the in-plane conduction in Nb$_{1-x}$B$_2$ are provided by the 2*p*( ) band, while carriers in MgB$_2$ were provided by the 2*p*( ) band as holes [19 and 20].

Furthermore, we examined the relationship between the sintering pressure and $T_c$ by fixing the sintering temperature at 1200 ˚C. Figure 5 shows the temperature dependence of the DC susceptibility of the samples prepared under various pressures. It is noted that the $T_c$ clearly increased with increasing the pressure. The lattice parameter $a$ was slightly elongated with raising the pressure, while there was almost no change in the lattice parameter $c$, as listed in Table 2.

3.2 Crystal structure and superconductivity in $Nb_{1-x}B_2$

A series of $Nb_{1-x}B_2$ for various x's was prepared at 5 GPa to examine the relationship between composition and superconductivity. In order to obtain a high-quality sample, attention was paid to the fact that the proper sintering temperature may depend on the composition. As previously mentioned, it was revealed that sintering at 1200 ˚C for 0.5–2 h was not enough to obtain a homogeneous sample for x = 0. Next, we prepared a sample for 2 h at 1300 ˚C, which is the highest temperature that can applied at present. The obtained sample showed sharper XRD peaks without an impurity peak. In the DC susceptibility of this sample, superconductivity was observed below 3 K with a very small volume fraction. Although the SEM image showed a small amount of impurities such as Nb, the sintering at 1300 ˚C gave a better than the sintering at 1200 ˚C for x 0. From the above results, we determined that the optimal conditions could be obtained by sintering at 1300 ˚C for 2 h for x = 0 and 0.04 and 1200 ˚C for 0.5 h for x = 0.08–0.52.

The temperature dependence of DC susceptibility of $Nb_{1-x}B_2$ is shown in Fig.6. A small amount of superconducting signal was detected for x = 0. However, we suspected that this superconductivity was introduced by some superconducting impurity such as unreacted Nb-metal and Nb-deficient- $Nb_{1-x}B_2$ that could not be removed in our synthetic process. Therefore, to examine the superconductivity of stoichiometric sample more carefully, we synthesized the sample with the Nb-rich nominal composition x = –0.05, which was expected to be close to a stoichiometric sample (refer to Table 2). This sample did not show any superconductivity in DC susceptibility above 2 K. On the other hand, a large superconducting volume fraction and a sharp transition at $T$ above 8.5 K were observed for 0.08 x < 0.24. The $T_c$ gradually increased with x and reached the maximum value of 9.2 K at x = 0.24, as indicated in Fig. 5(b). A further increase of x up to 0.48 led to a slight drop in the $T_c$.

The compositional dependence of the lattice parameters and $T_c$ are presented in Table 2 and plotted in Fig. 7. The lattice parameter *a* suddenly changed at x = 0.04 and monotonically decreased with increasing x. An abrupt change in the parameter *c* was also observed at x = 0.04, above which it slightly increased and then decreased with the maximum value at x=0.24. $Nb_{1-x}B_2$ was also prepared at 2 GPa over a wide range of x's, and the superconductivity was examined. The $T_c$'s of this series were about 1 K lower than those of the 5 GPa samples.

Furthermore, the samples with x ≥ 0.54 were synthesized and examined. Although no impurity peak was detected in the XRD patterns, the superconducting transition became broad and there was no more change in the lattice parameters.

3.3 Synthesis and characterization of $Ta_{1-x}B_2$ and $Mo_{1-x}B_2$

$TaB_2$ may be a good reference of $NbB_2$ because Ta and Nb are in the same group (5A) in the periodic table and have analogous chemical properties with slightly different metal ionic radii. Stoichiometric and metal-deficient $TaB_2$ with an $AlB_2$-type structure are known to exist [8-10 and 18]. Though superconductivity at 9.5 K in $TaB_2$ has been reported [6], the result has not been confirmed. In this study, $Ta_{1-x}B_2$ was prepared under high pressure in order to examine superconductivity with an expectation of raising the $T_c$ as in $Nb_{1-x}B_2$.

A series of $Ta_{1-x}B_2$ (x= 0–0.24) samples was synthesized at 5 GPa and 1200 °C for 0.5 h. No impurity peak was detected in the XRD patterns of any of the samples. Figure 8 shows the compositional dependence of lattice parameters *a* and *c*. Variation of both *a* and *c* is monotonic with the composition, which is in contrast to the case for $Nb_{1-x}B_2$. The present structure data are in good agreement with those reported by Itoh *et al.* [18] for the sample prepared under ambient pressure. A further increase in Ta-deficiency gave no more linear change in the lattice parameters, suggesting a phase formation limit. This indicates that high-pressure synthesis is not always effective for extending a solid-state range. The DC susceptibility of the obtained samples was carefully measured, but no superconductivity was found above 2 K at any composition.

The $Mo_{1-x}B_2$ system reported by Cooper *et al.* [11] was also examined using the high-pressure synthesis method. We prepared samples with compositions of $Mo_{1.0}B_{2.0}$ and $Mo_{0.84}B_{2.0}$ under 2–5 GPa at 1100–1200 °C. The XRD patterns of all the obtained samples indicated that the samples belong to a $Mo_2B_5$-type structure but not an $AlB_2$-type. No superconductivity signal was observed in the susceptibility. In this system, high pressure may induce the structural instability of $MoB_2$.

4. Discussion

4.1 Superconductivity in $Nb_{1-x}B_2$

The experimental results revealed that $Nb_{1-x}B_2$ prepared under high pressure was stable in a relatively wide range of x and exhibited superconductivity with a $T_c$ higher than 9 K. Here, the relation among the superconductivity, composition, and crystal structure in this system are discussed and compared with other metal diborides with $AlB_2$-type structure.

The first issue is which element is concerns non-stoichiometry, namely, whether Nb is deficient or whether there is an excess of B. As mentioned above, it is unlikely that there is an excess of boron in the lattice from the viewpoint of crystal chemistry. Moreover, our preliminary analysis of Reitveld refinement of the XRD pattern for $Nb_{0.84}B_2$ indicated an imperfect occupancy at Nb site. The Nb deficiency is also supported by a theoretical calculation. According to Oguchi [4], the formation energy of diborides decreases with increasing the number of valence electrons per atom, for example, from $ZrB_2$ to $MoB_2$. This decrease of formation energy may induce a metal defect [19]. A detailed structure analysis of $Nb_{1-x}B_2$ using neutron diffraction or single-crystal X-ray diffraction is necessary to clarify this problem.

The next issue that should be discussed is that the $T_c$ depended on the synthetic pressures. In order to confirm the pressure effect, $Nb_{0.84}B_2$ was prepared at an ambient pressure by the quartz-sealed method at 1050 °C for 10 h, starting with exactly the same raw materials as were used in the high-pressure synthesis. The value of $T_c = 5.3$ K for this sample was in good agreement with the $T_c$ reported by Akimitsu *et al.* (5.2 K) [14]. It is, thus, concluded that an appearance of superconductivity at 9.2 K is clearly owed to the pressure. Such a variation of the $T_c$ based on the synthetic method or heating conditions is well known in conventional intermetallic compound superconductors. For example, a $T_c$ (~18 K) of $Nb_3Si$, which is prepared by spattering and high-pressure methods is higher than those prepared under normal pressure [21]. In the case of $Nb_3Ge$, rapid cooling produces a higher $T_c$ [22]. These examples indicate that synthetic conditions including pressure are very important factors governing the $T_c$. The electronic structure of these materials is assumed to be sensitive to defects or randomness, and, as a result, a change in the density of state (DOS) at the Fermi level leads to variation in the $T_c$. The same explanation may be applicable for the change in $T_c$ with synthesis pressure in $Nb_{1-x}B_2$.

To realize a higher $T_c$ by changing a local structure, such as a micro strain, samples were synthesized with long sintering (~ 2 h), the slow cooling (~ 5 h), and the post-annealing (800 °C, 12 h), but these processes were all ineffective for changing the $T_c$. The $T_c$ in this system was found to depend mainly on the composition and the sintering pressure and not on the heating process.

The $T_c$ variation range (8.4 K ≤ $T_c$ ≤ 9.2 K) for 0.04 ≤ x ≤ 0.52 was not as wide as expected. We assume that the density of state (DOS) is almost flat near the Fermi energy and, thus, insensitive to a change in the number of valence electrons. A small change in $T_c$ might be assumed to indicate a kind of phase separation. However, it appears that there is a certain range of the Nb-deficiency in this system because the Nb-content determined by EPMA corresponds to the nominal composition of Nb, as listed in Table 2. Lattice parameters varied in synchrony with composition, as shown in Fig. 6.

In constant to the x-insensitive behavior of the $T_c$ for x > 0.04, the changes in the $T_c$ and the lattice parameters between x = 0 and 0.04 are remarkable. The strong correlation of these quantities suggests that the amount of band overlap, rather than the band filling, is a crucial parameter for $T_c$. The lattice parameters near x = 0.04 would be critical to gain a certain amount of DOS that is necessary for superconductivity.

To understand the physical properties, single crystal is necessary because the structural anisotropy of $Nb_{1-x}B_2$ must induce anisotropy in the physical properties. Although the growth of single crystal with a stoichiometric composition has been reported [22], its physical properties at low temperature were not sufficiently investigated. Single crystal with metal-deficient $NbB_2$ has not been reported yet. It may not be easy to obtain $Nb_{1-x}B_2$ single crystal with a $T_c$ above 9 K because high pressure is necessary for the growth of the crystal.

4.2 Comparison with other metal diborides $MB_2$

The composition and structure of $MB_2$ are finally discussed based on the present results together with the reported properties of $MB_2$ [6, 8–11, 13–15, 18, and 21–23].

In the present study, $Nb_{1-x}B_2$ and $Ta_{1-x}B_2$ were found to have a range of metal-deficiency in high-pressure synthesis, similarly to synthesis under ambient-pressure synthesis. According to the band calculation [3], the formation energies of $NbB_2$ and $TaB_2$ are relatively low, and, thus, the metal deficiency makes the materials stable. As the metal deficiency increased, the lattice parameter $a$ is reduced, and $c$ is expanded in both $Nb_{1-x}B_2$ and $Ta_{1-x}B_2$. However, these two compounds were quite different in the compositional dependence of the lattice parameters. $Nb_{1-x}B_2$ caused an abrupt change in the lattice parameters with increasing x, in contrast to the continuous change in the lattice parameter of $Ta_{1-x}B_2$. The lattice change must lead to a change in the electronic structure, and, thus, it would be necessary to perform a theoretical calculation on the compositional dependence of the band structure for $Nb_{1-x}B_2$ and $Ta_{1-x}B_2$.

Similar metal-deficient ranges were also reported in $VB_2$ [21] and $MoB_2$ [11], probably owing to the high valences of $V^{5+}$ and $Mo^{6+}$. However, there is no detailed study of the compositional dependence of the physical properties not even for samples prepared at ambient pressure. Although $Mo_{1-x}B_2$ was not prepared under high pressure in the present study, $Mo_{1-x}B_2$ could be another candidate for a superconductor.

In the case of $MgB_2$, a certain range of metal deficiency was reported [24–26]. With decreasing the nominal composition of Mg, the lattice parameter $a$ decreased, while $c$ increased, similarly to the case of $Nb_{1-x}B_2$ and $Ta_{1-x}B_2$. However, the absolute value of deficiency was much smaller than that of $Nb_{1-x}B_2$. In this case, metal deficiency cannot simply be discussed from the viewpoint of valence because the electronic structure of $MgB_2$ is different from that of other typical metal diborides with the $AlB_2$-type [2, 3].

On the other hand, there is no report about metal deficiency or non-stoichiometry in tetravalent metal diborides such as $TiB_2$, $ZrB_2$, and $HfB_2$. The absence of metal deficiency was also supported by the band calculation [3, 4]. Although they are very important materials for applications because of their high melting-temperature and hardness, they did not show superconductivity down to 0.26 K [13].

5. Summary

In the present study, the crystal structure and superconductivity of $Nb_{1-x}B_2$ prepared under high pressure were investigated. Although no clear superconductivity was observed down to 3 K for a stoichiometric sample of $NbB_2$, as the Nb-deficiency increased, the superconductivity appeared at x = 0.04 in $Nb_{1-x}B_2$, and the $AlB_2$-type structure remained. The abrupt increase in the $T_c$ at x = 0.04 coincided with the rapid change in the lattice parameters, which suggests that the amount of band overlap is a crucial parameter to control $T_c$. The transition temperature $T_c$ slightly rose and fell, as x increased, with the maximum value of 9.2 K at x = 0.24 for the samples were sintered at 5 GPa and 1200 ˚C. The value of $T_c$ = 9.2 K is the highest value for this material. The $T_c$ was found to rise with an increase in the synthesis pressure, in contrast to the insensitive compositional dependence of $T_c$. $Ta_{1-x}B_2$ and $Mo_{1-x}B_2$ were also synthesized under high pressure, but no superconductivity was observed for the Ta series, nor was any product with the $AlB_2$-type structure found for the Mo series.

Acknowledgement

We thank Dr. Oguchi of Hiroshima University for discussion of the band structure. This work was supported by the New Energy and Industrial Technology Development Organization (NEDO) as Collaborative Research and Development of Fundamental Technologies for Superconductivity Applications.

Table 1 Observed and calculated d-spacing for a sample with the nominal composition of $Nb_{0.84}B_2$ with $T_c$ = 9.1 K. Space group: P6/mmm (no.191) and lattice parameters: a = 3.0987(1) A and c= 3.3184(2) A

| h | k | l | $d_{obs.}$ /A | $d_{cal.}$ /A | Intensity |
|---|---|---|---|---|---|
| 0 | 0 | 1 | 3.316 | 3.318 | 40 |
| 1 | 0 | 0 | 2.681 | 2.683 | 91 |
| 1 | 0 | 1 | 2.085 | 2.086 | 100 |
| 0 | 0 | 2 | 1.659 | 1.659 | 5 |
| 1 | 1 | 0 | 1.549 | 1.549 | 14 |
| 1 | 0 | 2 | 1.411 | 1.411 | 14 |
| 1 | 1 | 1 | 1.404 | 1.403 | 15 |
| 2 | 0 | 0 | 1.341 | 1.341 | 5 |
| 2 | 0 | 1 | 1.244 | 1.243 | 9 |

Table 2 Synthetic conditions, lattice parameters and $T_c$ of $Nb_{1-x}B_2$ prepared under high pressure

(a) $Nb_{0.84}B_2$ prepared under various pressures

| Synthetic condition | a / A | c / A | V / A3 | a/c | $T_c$/K |
|---|---|---|---|---|---|
| 1.5 GPa, 1200 C, 0.5 h | 3.094(1) | 3.319(2) | 31.772 | 1.072 | 6.9 |
| 2.0 GPa, 1200 C, 0.5 h | 3.0972(4) | 3.316(1) | 31.809 | 1.071 | 8.3 |
| 3.5 GPa, 1200 C, 0.5 h | 3.099(1) | 3.320(2) | 31.884 | 1.071 | 8.8 |
| 5.0 GPa, 1200 C, 0.5 h | 3.0987(1) | 3.3184(2) | 31.863 | 1.0709 | 9.1 |

(b) Nb1-xB2 prepared 5 GPa, 1200 C, 0.5 h

| Nominal composition | a / A | c / A | V / A3 | a/c | Tc/K | Analyzed composition |
|---|---|---|---|---|---|---|
| 1.05a) | 3.1087(2) | 3.2705(4) | 31.606 | 1.052 | < 2 | Nb1.08B2.0 |
| 1a) | 3.1081(1) | 3.3278(3) | 31.667 | 1.0547 | 3 | -- |
| 0.96 a) | 3.1018(3) | 3.3156(8) | 31.899 | 1.0689 | 8.4 | -- |
| 0.92 | 3.1010(4) | 3.314(1) | 31.868 | 1.0687 | 9.0 | -- |
| 0.84 | 3.0987(1) | 3.3184(2) | 31.863 | 1.0709 | 9.1 | Nb0.76B2.0 |
| 0.76 | 3.0979(4) | 3.321(1) | 31.871 | 1.0720 | 9.2 | Nb0.68B2.0 |
| 0.68 | 3.0972(4) | 3.320(1) | 31.847 | 1.0719 | 9.0 | -- |
| 0.60 | 3.0962(5) | 3.319(1) | 31.817 | 1.0719 | 8.9 | -- |
| 0.52 | 3.0963(6) | 3.318(1) | 31.809 | 1.0716 | 8.7 | -- |

(a) Sintered at 1300 C for 2 h

(b) -- : did not measured or could not analyzed

Figure captions

Fig. 1 X-ray powder diffraction pattern of $Nb_{0.84}B_2$ prepared under 5 GPa at 1200 ˚C

Fig. 2 Images of cross-section of $Nb_{0.84}B_2$ (5 GPa, 1200 ˚C), upper: secondary electron and lower: reflected electron.

Fig. 3 Temperature dependence of (a) DC-susceptibility and (b) electrical resistivity for $Nb_{0.84}B_2$

Fig. 4 Temperature dependence of DC-susceptibility for $Nb_{0.84}B_2$ samples synthesized at various pressures

Fig. 5 Temperature dependence of dc-susceptibility for $Nb_{1-x}B_2$ (x = 0–0.24)

Fig. 6 Compositional dependence of lattice parameters and $T_c$ of $Nb_{1-x}B_2$ (x = 0–0.48)

Fig. 7 Compositional dependence of lattice parameters of $Ta_{1-x}B_2$ (x = 0–0.24)

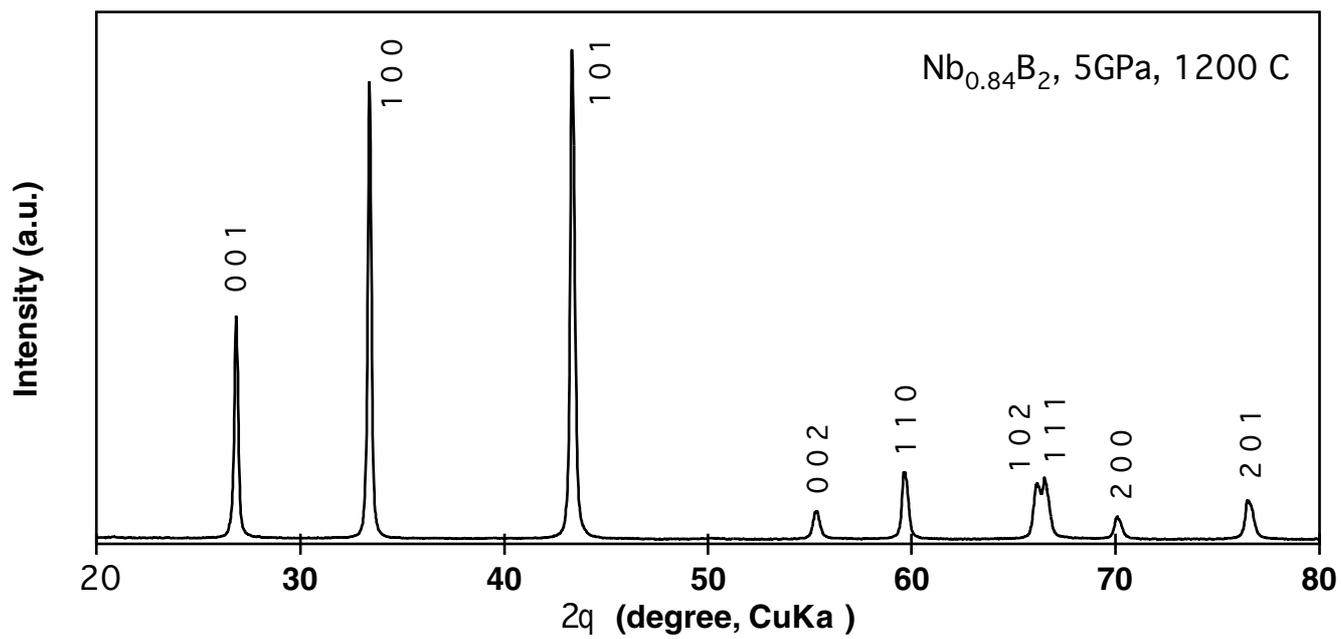

Fig.1 A.Yamamoto et al.

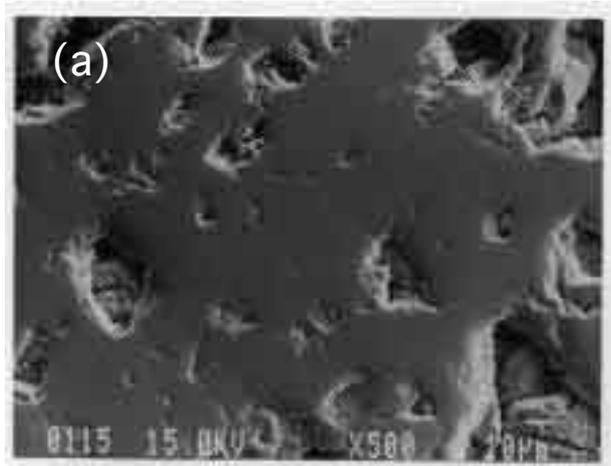

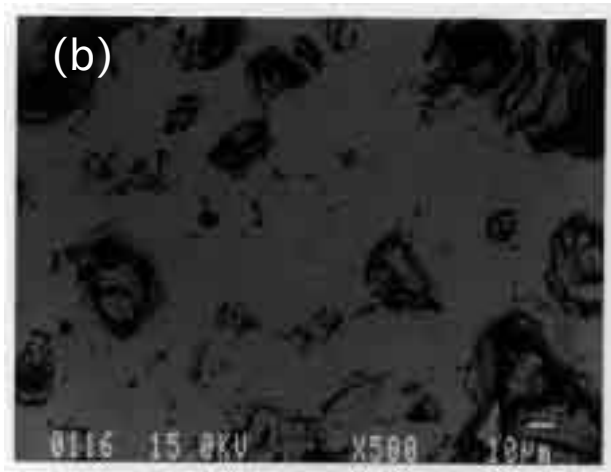

Fig. 2  A. YAMAMOTO et al.

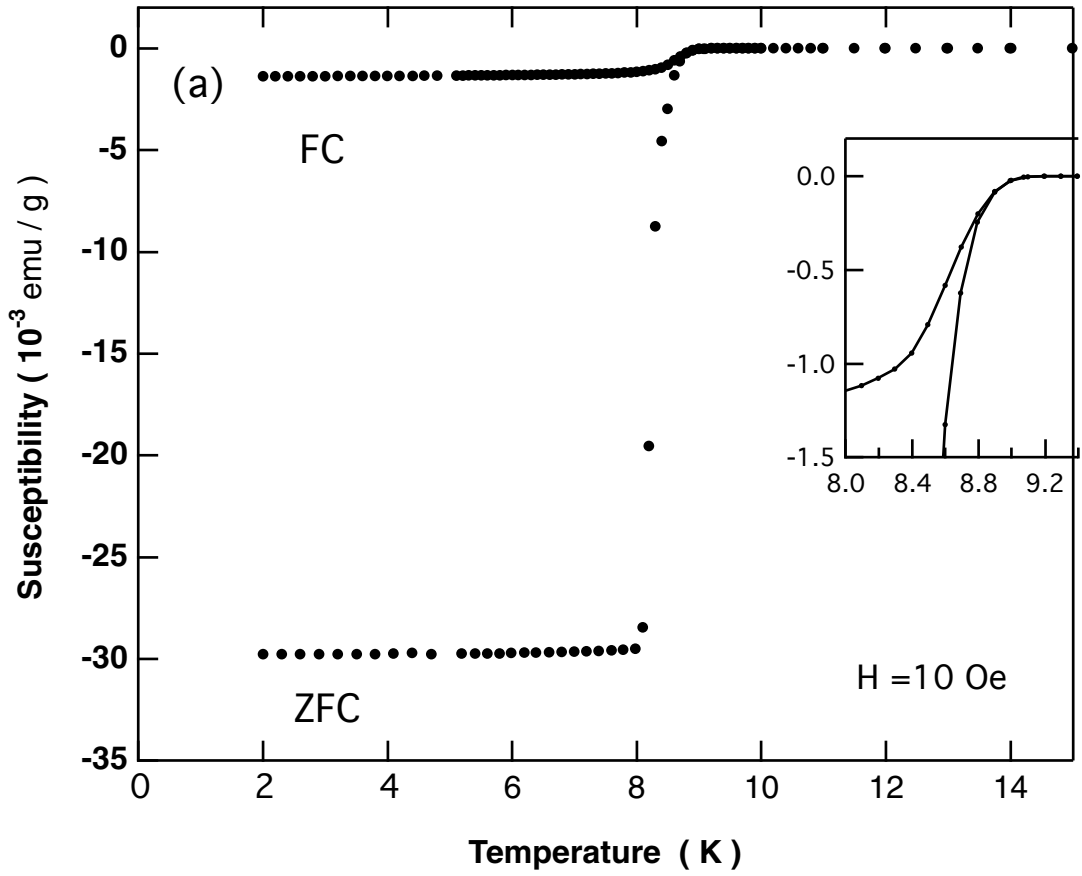

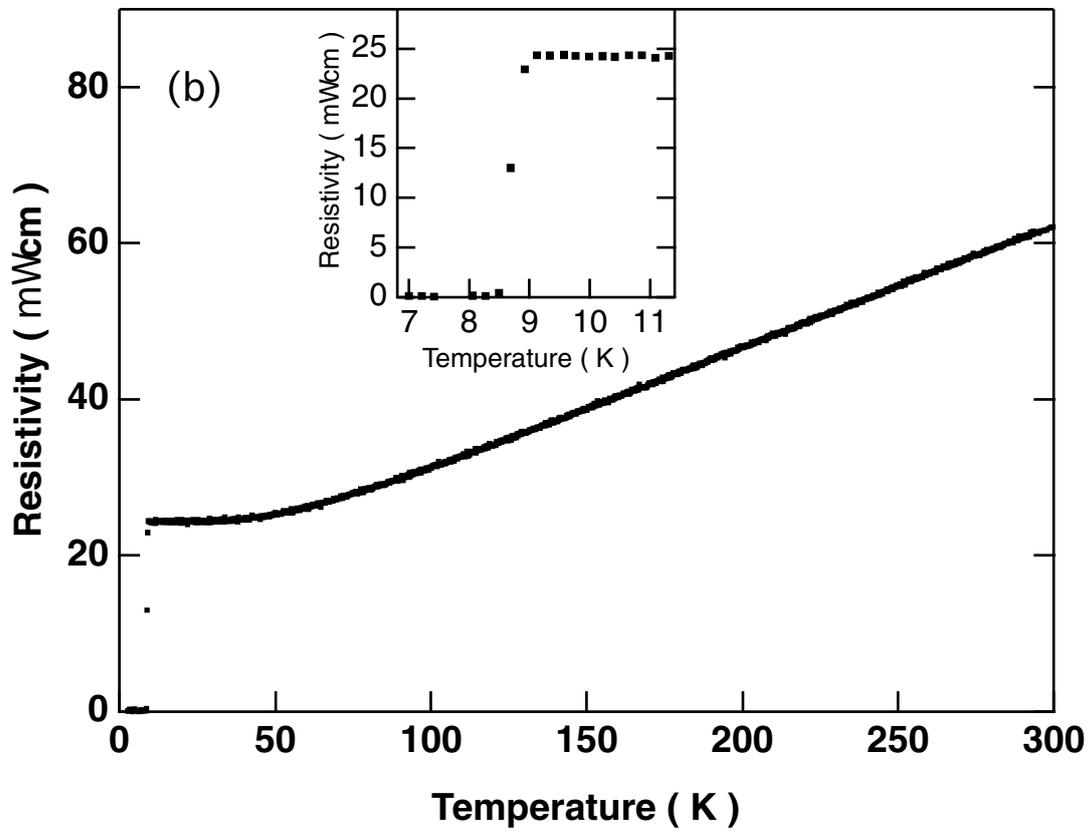

Fig. 3  A. Yamamoto et al.

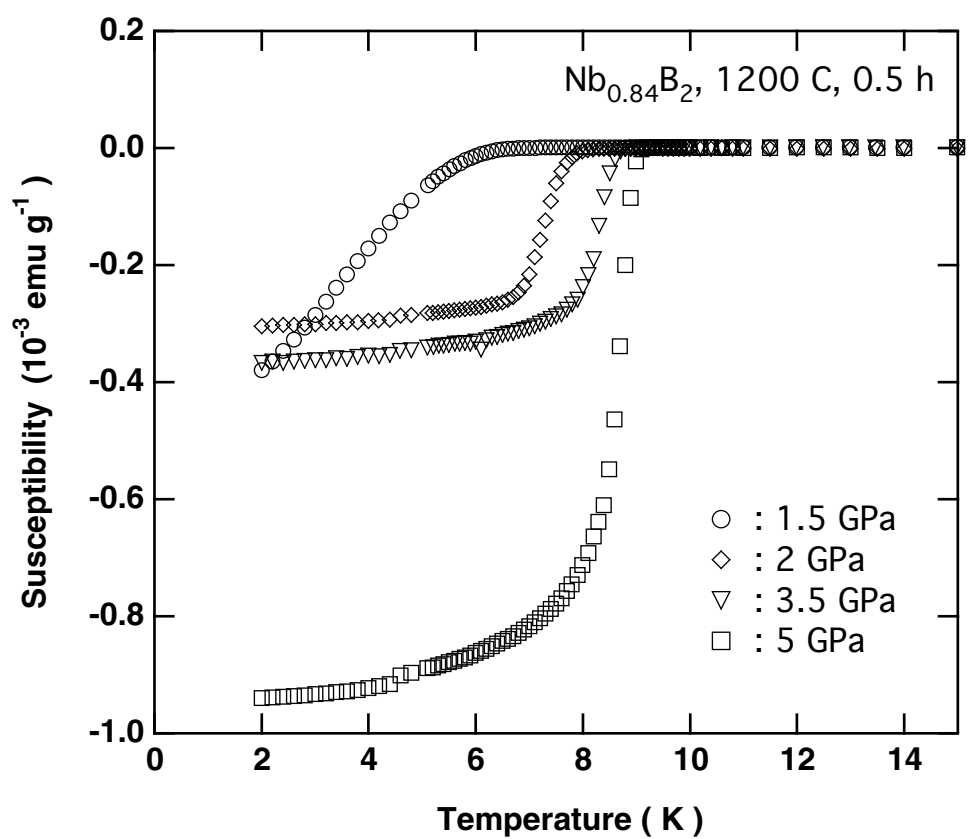

Fig. 4  A. Yamamoto et al.

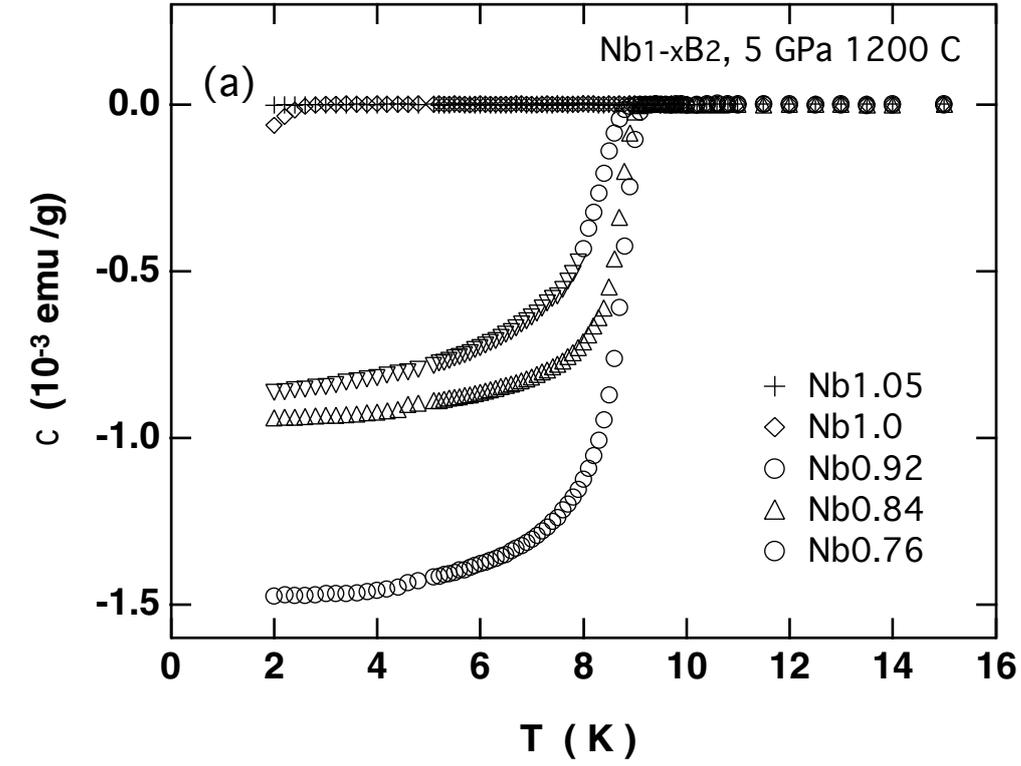
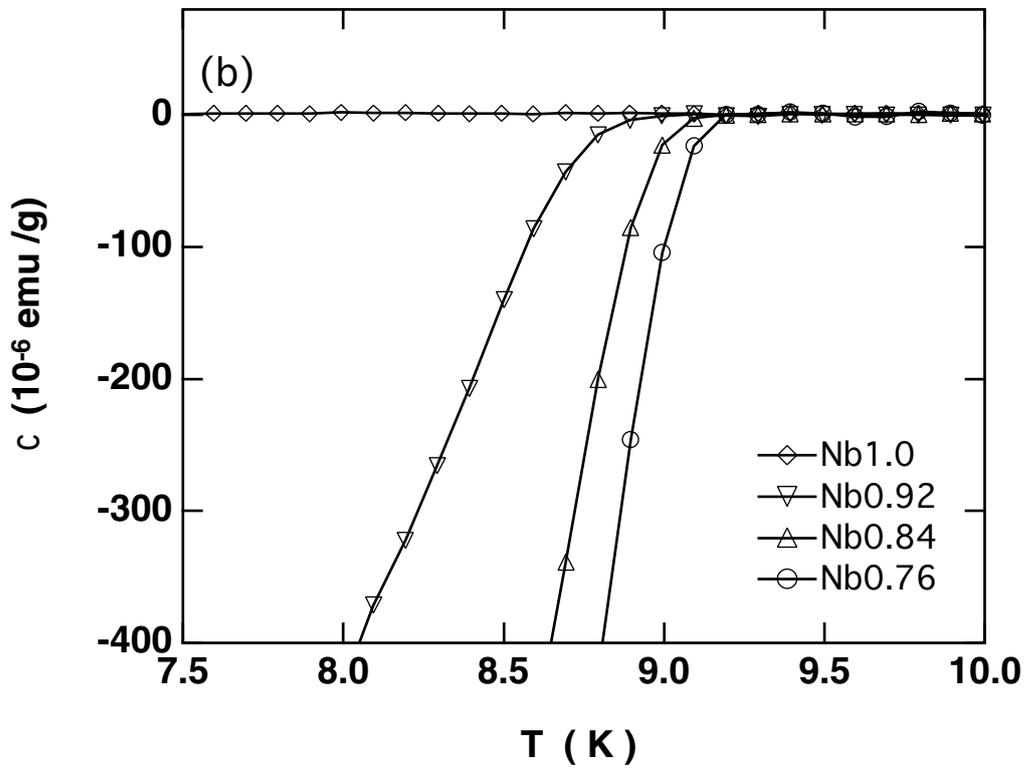

Fig. 5  A. Yamamoto et al.

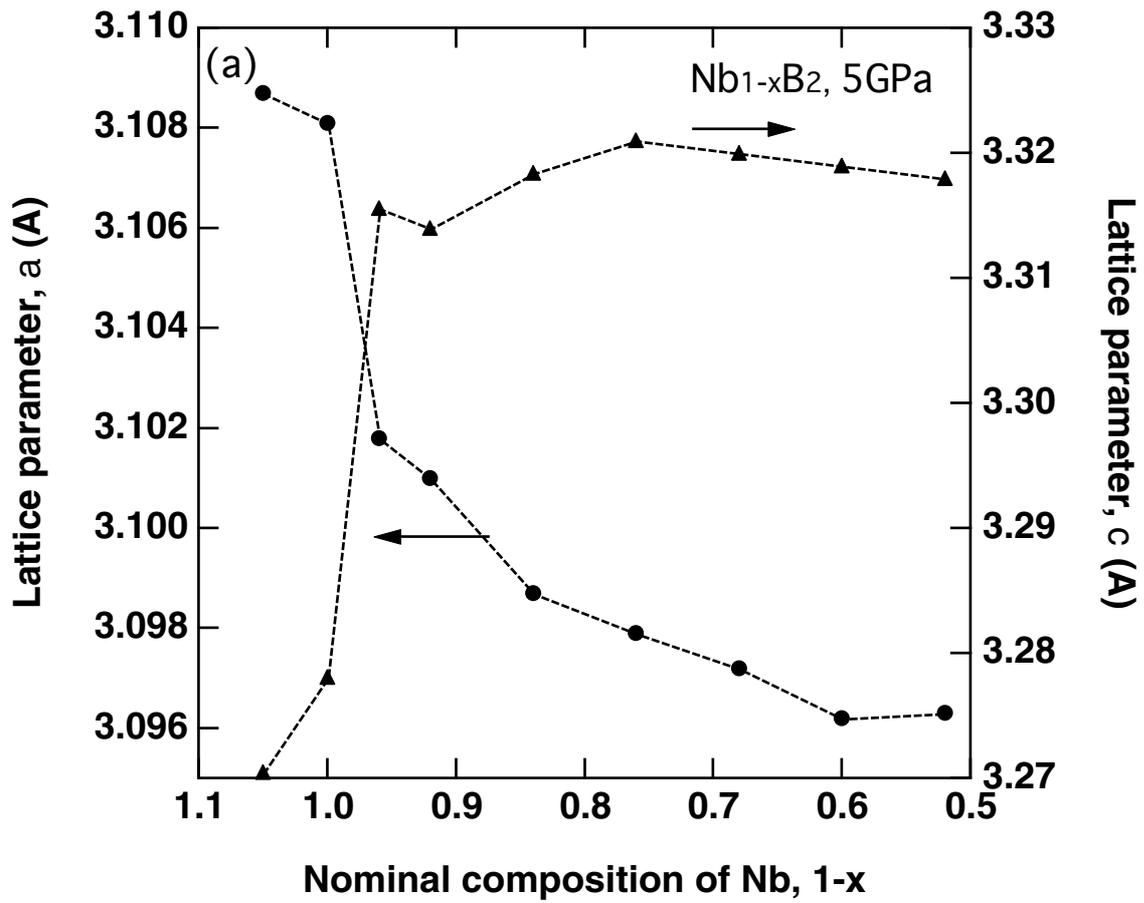

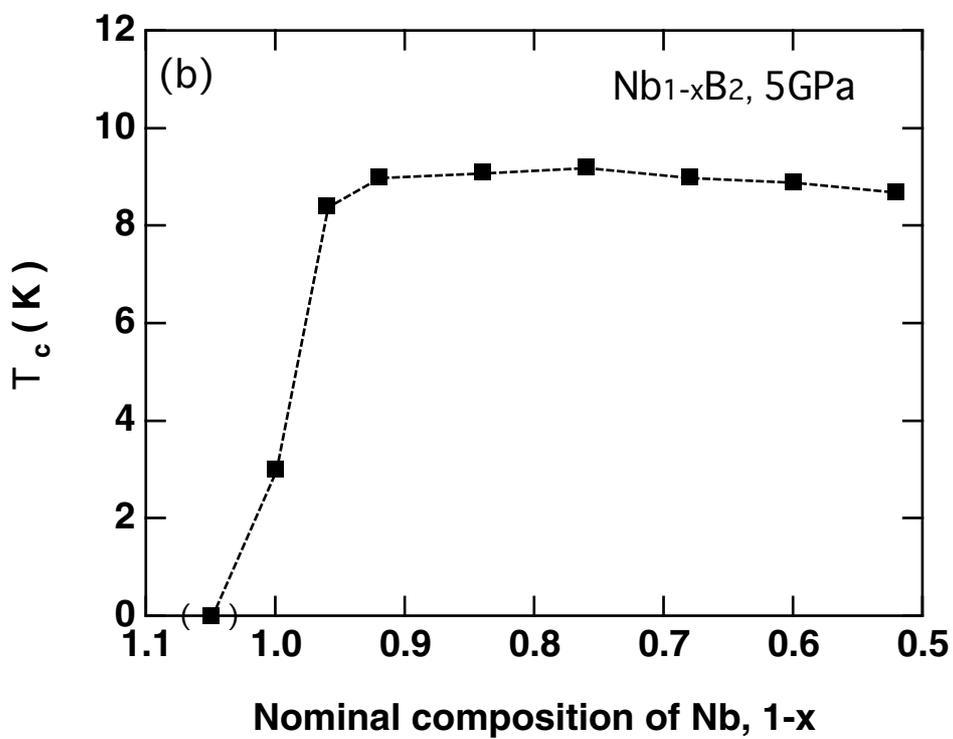

Fig. 6 A. Yamamoto et al.

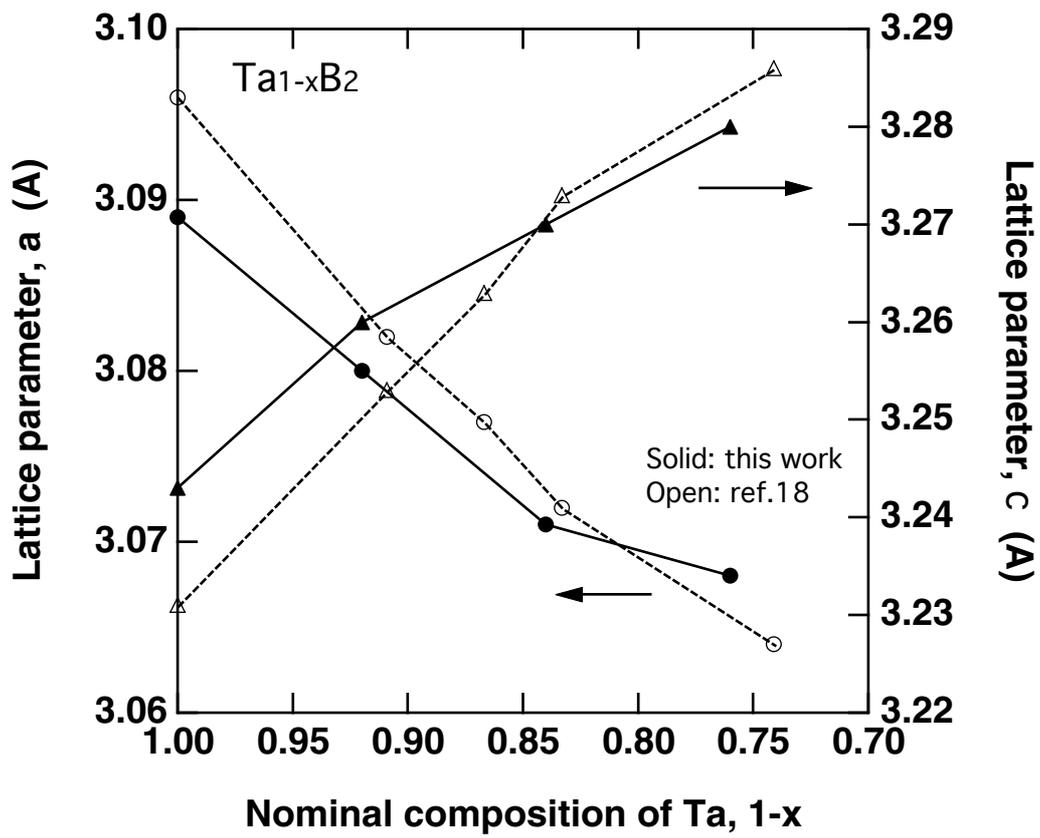

Fig. 7  A. Yamamoto et al.